# Out-of-plane Ionicity versus In-plane Covalency Interplay in High-$T_c$ Superconducting Oxides

T. Guerfi[1]

Department of Physics, Faculty of Science, M'Hamed Bougara University, Boumerdes, Algeria

**ABSTRACT**

It seems that the remarkable properties of the high temperature superconducting oxides, especially the Insulator-Metal Transition (IMT) and the Metal-Superconductor Transition (MST) both originate from the competition (interplay) between ionic versus in-plane covalence nature of bonds in these materials. As a result of this competition, the microscopic order parameter, that is firmly identified to be the local field estimated from the ionic polarization at the sub-unit level (one half of the unit cell), shows a strong temperature as well as chemical doping dependence. While the out-of-plane ionicity is responsible for the interlayer charge transfer (electrons) that reduces it leading to IMT, the in-plane covalency is responsible for the in-plane intersite transfer of charge (holes) that increases the ionicity and leads to MST.

This interplay of charge transfer, driven by the out-of-plane ionicity and the in-plane covalency, leads at the critical temperature $T_c$ to a local field avalanche (ionic polarization catastrophe at the sub-unit cell level) that triggers an out-of-plane correlation between the local polarization fields at the unit cell and drives the compound into the superconducting state. The asymmetry of the free charge carrier density breaks locally the inversion symmetry of the order parameter leading to an unconventional pairing mechanism characterized by the CPT transformation where C is the charge conjugation symmetry, P is the parity symmetry and T is the time reversal symmetry.

It is established that the out-of-plane ionicity versus in-plane covalency is a necessary condition for the occurrence of both IMT and MST. The underlined microscopic mechanism provides a route to a possible unified description of high-$T_c$ superconductivity.

---

[1] Corresponding author : tarek.guerfi@gmail.com
Tel: +213555237358

## 1. INTRODUCTION

The ionic model has provided the basis for our understanding of the very wide range of physical phenomena displayed by ionic crystals [1, 2]. However an unusual feature of ionic versus covalence nature of bonds was the origin behind the peculiar electrical and structural behaviors of some ionic crystals known to display a large values of a spontaneous polarization [3]. It is established that a mixed ionic-covalent B-O bonds is a necessary condition for the occurrence of a spontaneous polarization in $ABO_3$ type of ferroelectric crystals [4].

The High Temperature Superconductors (HTS) are ionic crystals with the parent compound to be charge transfer insulator [5]. In fact, all cuprate high-Tc superconductors are stacks of alternating anionic $CuO_2$ layers and block layers which are in gross cationic. The stacking direction is commonly taken as the crystallographic c-direction, and the crystal symmetry is generally derived from a simple tetragonal or a body centered tetragonal (b.c.t) symmetry, although it is in fact often orthorhombic due to distortions or due to vacancy ordering. The key element shared by all the structures of cuprates is the $CuO_2$ plane which has particular feature; its in-plane strong covalent character of bonds. The physical properties and the superconducting $T_c$ are strongly influenced by the density of charge carriers in the $CuO_2$ plane, which is regulated by variation of the charge-reservoir block composition. However, the maximum $T_c$ varies widely between crystals sharing the same basic $CuO_2$ in-plane electronic structure, by up to a factor of 10 at the same hole density in monolayer cuprates [6]. These variations, as it has been mentioned obviously [7, 8], cannot be due to the doping dependence of the $CuO_2$ in-plane electronic structure, and it has been hypothesized for a long time that there must be a key out-of-plane influence [6] that controls the basic electronic structure of cuprates.

Also, the role of the lattice in the high temperature superconducting materials was demonstrated by Angle Resolved Photoemission Spectroscopy (ARPES) [9] and X-ray Absorption Spectroscopy (XAS) which revealed a local displacement of oxygen as carriers are doped indicating an intimate relation between the lattice effects and the pairing mechanism [10]. Moreover, isotope and strain (pressure) effects are considered as direct experiments that establish the role of the lattice in the HTS [11].

Through a competition between the ionic nature of bonds versus in-plane covalence nature of bonds mainly in the $CuO_2$ plane (the cuprates case), the local field that reflects interlayer polarization estimated from the ionic polarization at the sub-unit cell level (one half of the unit cell) along the *c*-axis which is the polar axis, is the microscopic order parameter responsible for the in-plane and the out-of-plane physics of high-$T_c$ superconductors[12, 13] .

By means of chemical doping which consists in most cases in adding oxygen to the parent insulator compound, the interlayer local field can be reversed with respect to the parent undoped case especially between the anionic $CuO_2$ and the adjacent cationic layer. A possible arrangement of the electronic structure that lowers the energy of the system consists in an interlayer transfer of electrons from the anionic $CuO_2$ layer to the adjacent cationic layers reducing the ionicity of the layers and cancelling the resulting local electric field which otherwise would be present. This mechanism is the origin behind the observed Insulator-Metal phase Transition (IMT).

Although the IMT reduces the iconicity and causes the local field (the order parameter) to vanish, however, the strong in-plane covalency of $CuO_2$ induces an in-plane intersite transfer of holes form oxygen to copper; this transfer of charge increases the iconicity and causes the local field to reemerge again. At a given critical temperature, this competition between ionicity and covalency leads to a sort of local field avalanche that triggers an out-of-plane correlation between the local polarizations fields at the unit cell level driving the metal into the superconducting state.

## 2. CHEMICAL DOPING AND TEMPERATURE DEPENDENCE OF THE MICROSCOPIC ORDER PARAMETER IN YBA$_2$CU$_3$O$_{7-\delta}$

The HTS are ionic crystals with the parent compound to be the charge transfer Mott insulator [5]. The oxygen *p*-level and the cu *d*-level form hybridized bonding and anti-bonding orbitals states and the gap lies between the filled *p*-like of O band and the empty Cu *d*-like band [14]. In the purely ionic model for the HTS compounds, nominal valence charges are taken in order to compute the ionic polarization at the sub-unit cell level (the order parameter) in these materials. It is noted that these materials are centro-symmetric, so, the total ionic polarization at the unit cell level vanishes at any given temperature and there would be no external electric effects. However, the value of the local electric field that acts at the site of an ion is significantly different from the value of the macroscopic electric field. The reason of that is that by definition the macroscopic field is the field which is averaged over large number of dipoles, however the local field which acts on a particular atom is influenced by the nearest surrounding, and therefore can deviate from the average field.

The most prominent evidence of the strong temperature as well as the oxygen dependence of the local ionic polarization field (the order parameter) has been found in YBa$_2$Cu$_3$O$_{7-\delta}$. The crystal structure of YBa$_2$Cu$_3$O$_6$ and YBa$_2$Cu$_3$O$_{7-\delta}$ is considered as juxtaposition along z-axis of successive layers including the anionic CuO$_2$ layer which is characterized by strong in-plane hybridization. The resulting ionic polarization at the sub-unit cell (one half of the unit cell) is the order parameter, see figure. 1. The normal and the superconducting properties of this oxide are strongly dependent on oxygen doping as well as on temperature. Experimentally, in the range of oxygen content $0 \leq \delta \leq 0.65$, this compound has an orthorhombic crystal structure with a metallic behaviour at a normal state and superconducting behaviour at a critical temperature which decreases as the oxygen content decreases [15]. At an oxygen content corresponding to $\delta = 0.65$, this oxide undergoes a simultaneous Insulator-Metal Transition (IMT) and an orthorhombic to tetragonal structural phase transition [15]. Lowering more the oxygen content $0.65 \leq \delta \leq 1$, the material remains insulator in the tetragonal phase.

For an ideal ionic crystal the point charge model would apply. In the crystalline system, the macroscopic ionic polarization $P_{CM}$ is defined as the sum of dipoles moments in a given cell divided by the cell volume,

$$P_{CM} = \frac{1}{\Omega} \sum q_i r_i \qquad (1)$$

$\Omega$ is the unit cell volume,

$q_i$ are the formal ionic charges located at the atomic positions $r_i$.

The formal charges $q_i$ assigned for the different constituents ions are +3 for Y, +2 for Ba, -2 for O and assigning a formal charge for Cu balancing exactly the negative total valence charge of the oxygen. In case of YBa$_2$Cu$_3$O$_{6.9}$, the formal valence charge assigned for Copper is +2.2667. The positions of the ions $r_i$ have been chosen on the basis of high resolution neutron diffraction crystallographic data from reference [16]. The oxygen content dependence of the order parameter estimated from the ionic polarization computed on one half of unit cell in YBa$_2$Cu$_3$O$_{7-\delta}$ at 5 K is displayed in figure. 2. The order parameter $P(6+x)$ shows almost a linear behavior for all oxygen content, however it vanishes for oxygen content corresponding to $\delta = 0.65$ and its sign is reversed for $0 \leq \delta \leq 0.65$ oxygen content range. The maximum value of the order parameter, at 5 K, was obtained for optimally oxygenated samples.

The oxygen dependence of the order parameter (the local ionic polarization field) correlates very well with the oxygen dependence of YBa$_2$Cu$_3$O$_{7-\delta}$ experimental properties. It is noteworthy that YBa$_2$Cu$_3$O$_{7-\delta}$ undergoes an IMT at oxygen content $\delta = 0.65$ while the order parameter (the local ionic polarization field) vanishes at the same oxygen content $\delta = 0.65$.

By means of the chemical doping which consists in almost cases in adding oxygen to the parent undoped compound, the local ionic polarization field (the ionic polarization at sub-unit cell level) can be reversed with respect to the parent undoped case especially between the anionic $CuO_2$ and the adjacent cationic layer (see figure. 1 and figure. 2). For $0.65 \leq \delta \leq 1$ oxygen content that corresponds to the insulating state of $YBa_2Cu_3O_{7-\delta}$, the resulting local ionic field at the sub-unit cell level points always from the anionic $CuO_2$ layer to the cationic layer as it is the case in the parent undoped material $YBa_2Cu_3O_6$. In fact, in the range of $0 \leq \delta \leq 0.65$ oxygen content, the local polarization field is reversed as it is illustrated in figure 3. Every layer will experience the resulting electric field and in order to minimize these polarizations, each layer's constituent ions will move along the z-axis in the direction imposed by the resulting electric field and the signs of their charges, until they reach their equilibrium (real) positions, see figure. 4. So the apical oxygen O(4) and the Ba ions will move in opposite directions along the z-axis relatively to the parent undoped case. The apical oxygen O(4) tends to approache the anionic $CuO_2$ layer conversely to Ba. But mainly because of the strong in-plane covalency, the copper Cu(2) and the Oxygen O(2, 3) do not feel much the field as no significant displacements is observed, see figure. 4. The other important effect of this local field is to tilt all the bands; but mainly because of the strong covalent character of bonds of the $CuO_2$ layer, the bands that correspond to the cationic layer are lowered relative to those of the anionic $CuO_2$ layer compared to the parent undoped case of $YBa_2Cu_3O_6$, causing a possible arrangement of the electronic structure that consists in an interlayer transfer of electrons from the anionic layer $CuO_2$ to the adjacent cationic layer BaO. This mechanism reduces the ionicity of the layers and removes the resulting local electric field driving a hole doping of the p-like band of O. It is noteworthy that a similar mechanism involving charge transfer of electrons stabilizing the structure of polar oxide surfaces has been observed to also lead to 2D metallic surface state [17]. Accordingly, in all the range $0 \leq \delta \leq 0.65$ of oxygen content, due to IMT the oxide of $YBa_2Cu_3O_{7-\delta}$ is metallic, otherwise it is insulator.

For a single phase specimen of optimally oxygenated sample $YBa_2Cu_3O_{6.9}$ ($x=0.9$), a complete diffraction patterns using high resolution neutron powder diffraction taken at 300 K, 182 K, 101 K, 92 K, 82 K and 16 K have been analyzed by profile fitting methods [18] and the refinement procedures were processed exactly in the same way for all data sets in order to guarantee a reliable comparability of the results obtained for the different temperatures, so that any possible systematic errors occurring either in the experiment or in the data analysis were identical for all temperatures.

For the atomic position, we have used the crystallographic data taken from [18] and by assuming a formal valence charges for the different constituent ions, the point charge model would always apply. The temperature dependence of the local ionic polarization field (the order parameter) in $YBa_2Cu_3O_{6.9}$ is displayed in figure. 6. The compound $YBa_2Cu_3O_{6.9}$ exhibits a local ionic polarization of 0.1959 C/m$^2$ along the *c*-axis (the polar axis) at 300 K and of 0.2100 C/m$^2$ at 92 K, which is the superconducting transition temperature. The main feature of the temperature dependence of this local polarization is its step like increase at the superconducting transition temperature which can be qualified as a local ionic polarization catastrophe at $T_c$.

In ferroelectric oxides, the Born effective charge *Z\**, which relates the change of polarization *ΔP* with the displacement *u* as $\Delta P = Z^* eu / V_{cell}$ is enhanced by a strong orbital hybridization as compared to the formal valence state of ions. In these oxides the covalency induces an electronic polarization that points in the same direction as the ionic polarization doubling the Born effective charge in many cases [19, 20]. However, As it was pointed out by Egami [14], even though the doped cuprates are metallic, the ionic charges are not fully screened, and anions and cations approximately maintain their valences. Their mobile carrier density is low, and screening is ineffective at short distances. With displacement of copper toward oxygen, covalency induces an in-plane hole transfer from oxygen to copper. This time although, the electronic polarization points in the opposite direction to the ionic polarization, it increases the local charge of both ions (the ionicity increases) so that the vanishing local polarization field (the order parameter) reemerges again. There is a large volume of literature that shows coupling of superconductivity to the lattice and phonon [9, 21-23]. In particular, the in-plane Cu−O Longitudinal Optical (LO) bond-stretching phonon mode was observed by neutron inelastic scattering to show a strong softening with doping near the zone boundary along the Cu−O bond direction [21, 24, 25]. This mode induces charge transfer between Cu and O, and thus couples strongly to the charge [22, 26, 27]. Also the LO phonon dispersion observed by inelastic neutron scattering measurements on $YBa_2Cu_3O_{7-\delta}$ (YBCO) suggests that the frequencies of the LO phonons are strongly softened with doping near the zone boundary [28].

The displacements, in opposite directions along the *z*-axis of the center of positive and negative ions $G^+$, $G^-$ in the sub-unit cell as function of the temperature, illustrated in the figure. 7, are the main signature of the emergent local field (the order parameter). But in the same time, this local field may always be removed and the ionicity is reduced, by an interlayer transfer of electrons form the anionic $CuO_2$ layer to the cationic layer. However, this mechanism leads to an increase in the density of holes in the *p*-like band of O and in turn it causes again an intersite hole transfer toward copper. At a given critical temperature, this interplay between ionicity versus in-plane covalency causes an ionic sub-unit cell polarization catastrophe or a local field avalanche. The signature of this local field avalanche is the discontinuity in the shift of the position of the center of positive and negative ions $G^+$, $G^-$ respectively in opposite direction along the z-axis, observed at the superconducting transition temperature (figure. 6), but the shifts $G^+$ and $G^-$ are limited to finite displacements because of the coupling at the unit cell level of these local fields. In other words, the local ionic polarization field catastrophe at the critical temperature $T_c$ triggers the correlation out-of-plane between the local fields at the unit cell level and superconductivity is onset.

The local field avalanche (the ionic polarization catastrophe) can be considered as the origin behind many structural instabilities and anomalies that have been observed by many research groups using different experimental methods [18, 29-36] in $YBa_2Cu_3O_{6.9}$ as well as other HTS near the critical temperature $T_c$.

## 3. UNCONVENTIONAL PAIRING MECHANISM

The local polarization field (the order parameter) at the unit cell level vanishes at any given temperature in the normal state. However, because of the interplay between the ionicity and the covalency the local field tends to become infinite at the avalanche state, but the coupling at the unit cell level of the local fields stopped the field avalanche and at the same time a perfect out-of-plane correlation between these local fields at the unit cell level is established. The mirror plane which is the yttrium plane divides the unit cell into two coupled right and left sub-units cell. So, any decrease/increase of the local field in the right/left sub-unit cell is correlated by a local field increase/decrease in the left/right sub-unit cell. However in the avalanche (superconducting) state, the asymmetry of the free charge carrier density will break the inversion symmetry of the microscopic order parameter (figure. 8). As a consequence of this broken symmetry, the local field is no longer vanishing at the unit cell level leading to an unconventional pairing between the real current and the polarization current, or particle antiparticle pairing characterized by CPT transformation, where C is the charge conjugation symmetry, P is the parity symmetry and T is the time reversal symmetry.

## 4. CONCLUSION

It seems that the remarkable properties of the high temperature superconducting oxides (HTS), especially the Insulator-Metal Transition (IMT) and the Metal-Superconductor Transition (MST) both originate from the competition (interplay) between the ionic versus the in-plane covalence nature of bonds in these materials. The microscopic order parameter that has been identified to be the local polarization field estimated from the ionic polarization at the sub-unit cell (one half of the unit cell) shows a strong temperature as well as the chemical doping dependence, as a result of this competition.

It is showed that with chemical doping the local ionic polarization field can be reversed. An arrangement of the electronic structure that consist in an interlayer transfer of charge, between the anionic $CuO_2$ layer and the adjacent cationic layer, removes the resulting local polarization field and induces an IMT. However, the in-plane covalency induces also an intersite transfer of charge that increases the ionicity and causes the vanishing local polarization field to reemerge again. This interplay (competition) between ionicity-induced interlayer electron transfer and in-plane covalency-induced intersite hole transfer leads to a local field avalanche (a local ionic polarization catastrophe) at a given critical temperature $T_c$. This avalanche effect triggers an out-of-plane correlation between the local fields at the unit cell level driving the oxide into the superconducting state. The asymmetry or the free charge carriers density break locally the inversion symmetry of the order parameter leading to an unconventional pairing mechanism between the conduction current and the polarization current that is characterized by the CPT symmetry transformation.

**ACKNOWLEDGEMENT**

The author would like to thank Dr S. Toumi for her assistance in the correction of the present manuscript and for numeral fruitful discussions.

## FIGURE CAPTIONS

**Figure. 1** Crystal layered structure of $YBa_2Cu_3O_6$ and $YBa_2Cu_3O_{7-\delta}$ considered as juxtaposition along z-axis of successive layers including the anionic $CuO_2$ layer which is characterized by strong in-plane hybridization. The resulting ionic polarization at the sub-unit cell (one half of the unit cell) is the order parameter. M denotes the mirror plane which is the yttrium plane that divides the unit cell into two sub-unit cell

**Figure. 2** Oxygen content dependence of the order parameter (the ionic sub-unit cell polarization) in $YBa_2Cu_3O_{7-\delta}$ [13]. It is noteworthy that while the microscopic order parameter vanishes for $YBa_2Cu_3O_{6.35}$ ($\delta=0.65$), the oxide $YBa_2Cu_3O_{6.35}$ undergoes an insulator-metal phase transition at the same oxygen content

**Figure. 3** Z coordinate of the center of cations $G^+$ and the center of anions $G^-$ along with z coordinate of anions and cations in the sub-unit cell, in the undoped $YBa_2Cu_3O_6$ and in the doped $YBa_2Cu_3O_{6.95}$

**Figure. 4** Relative displacements $(z*c - z_r*c_r)/(z_r*c_r)$ along z axis, as function of oxygen doping, of Ba, O(4), Cu(2) and O(2, 3). $z_r$, $c_r$ is the z coordinate of the corresponding ion and the c-axis parameter respectively in the parent undoped $YBa_2Cu_3O_6$. Ba and O(4) of the cationic layer move in opposite directions in strong response to the resulting local field however Cu(2) and O(2, 3) of the anionic layer do not feel much the field as no significant relative displacement is observed ; note also that their displacements are in the same direction

**Figure. 5** Charge transfers inside the unit cell of $YBa_2Cu_3O_{7-\delta}$. By means of chemical doping, the local field can be inversed with respect to the parent undoped material $YBa_2Cu_3O_6$; while a charge transfer (electrons) from the anionic $CuO_2$ layer to the cationic BaO layer reduces the ionicity of the layers and removes the local field leading to a metallic sate of the $CuO_2$ layer, the in-plane covalency of $CuO_2$ induces a transfer of charge (holes) from the in-plane oxygen O(2, 3) to the in-plane copper Cu(2) increasing the ionicity and rising up the vanishing field . This interplay causes a local field avalanche at a given critical temperature $T_c$ that drives the metal into the superconductivity state

**Figure. 6** The temperature dependence of the order parameter in $YBa_2Cu_3O_{6.9}$ [13]. A step-like increase of the order parameter is observed at the critical temperature 92 K; the ionic polarization catastrophe or a local field avalanche that results from the competition between ionicity and the in-plane covalency in the oxide

**Figure. 7** Temperature dependence of the relative displacement of the center of cations $G^+$ and the center of anions $G^-$ normalized to 300 K. A discontinuity in the shift of the $G^-$ and $G^+$ is observed at the critical temperature $T_c$ which is the signature of the local field avalanche

**Figure.8**. (a) Above $T_c$ the Insulator-Metal Transition (IMT) reduces the ionicity of the compound, therefore it cancels the order parameter. b) Due to in-plane covalency-induced intersite charge transfer, the vanishing order parameter reemerges again. This competition competition between ionicity versus covalency causes an ionic polarization catasrophe at $T_c$ (local field avalanche) that tiggers an out-of-plane correlation between the local fields at the unit cell level. (c) The mirror (inversion) symmetry of the microscopic order parameter is localy broken by the asymmetry of the free charge carriers density. (d) Uncoventionnal pairing mechanism between the real current and the polarization current characterized by CPT symmetry that results from the broken symmetry of the order parameter

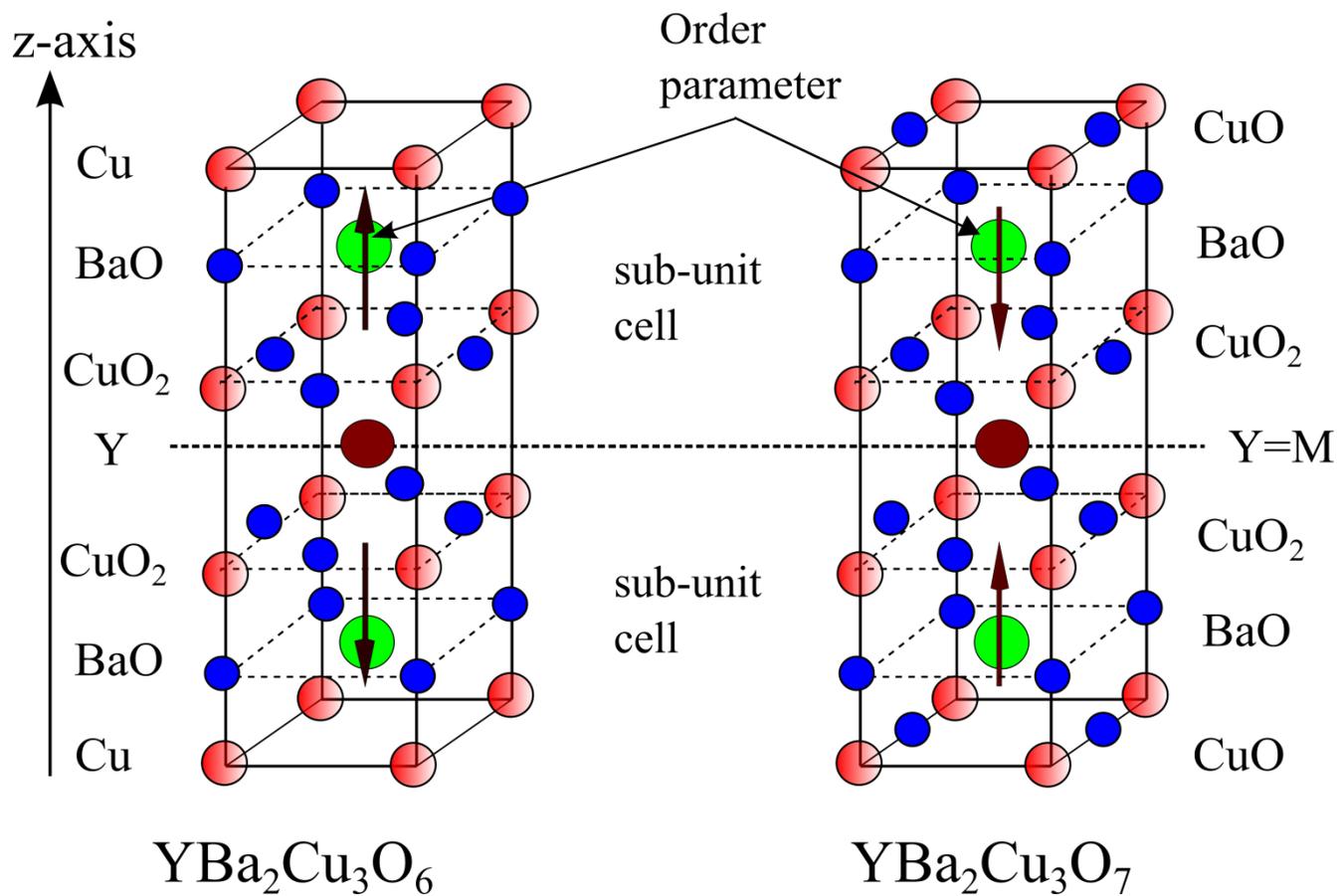

Figure. 1

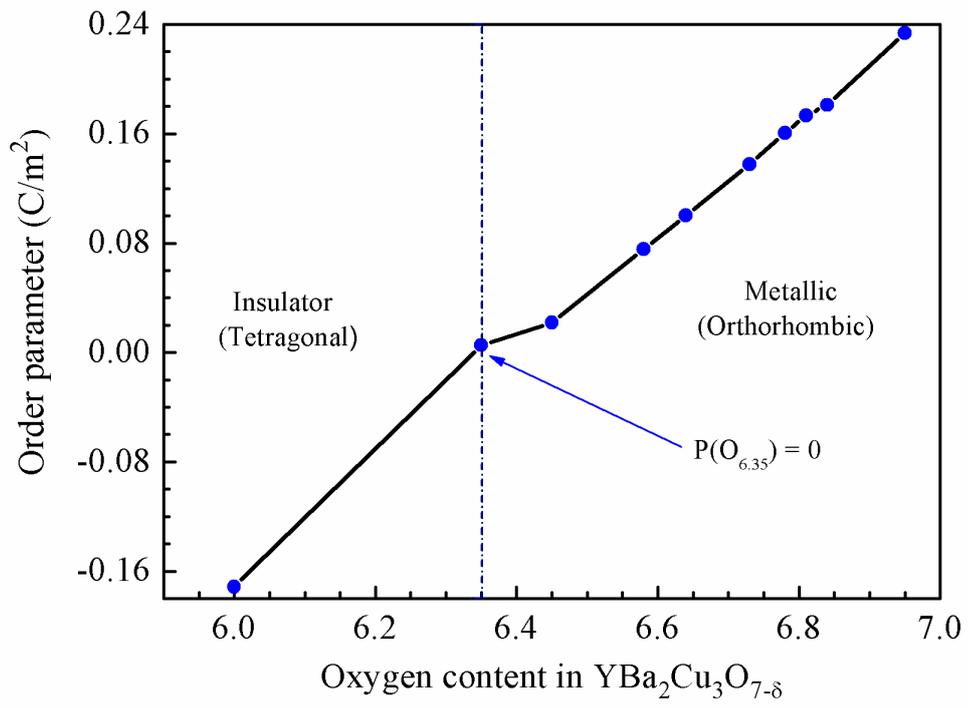

**Figure.2**

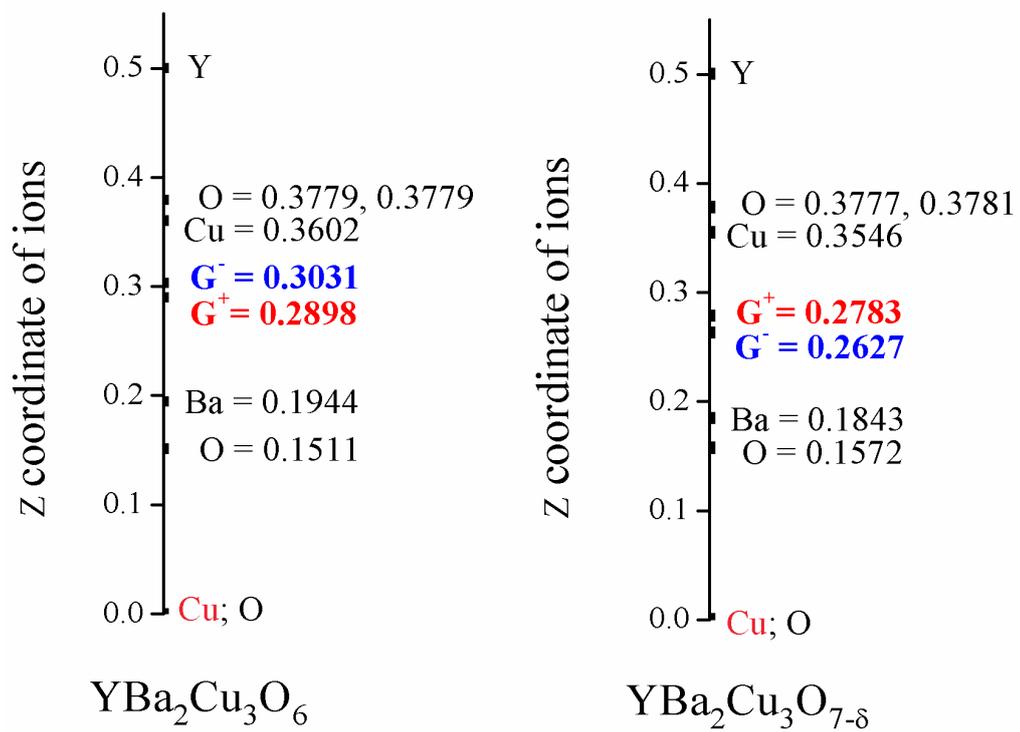

**Figure. 3**

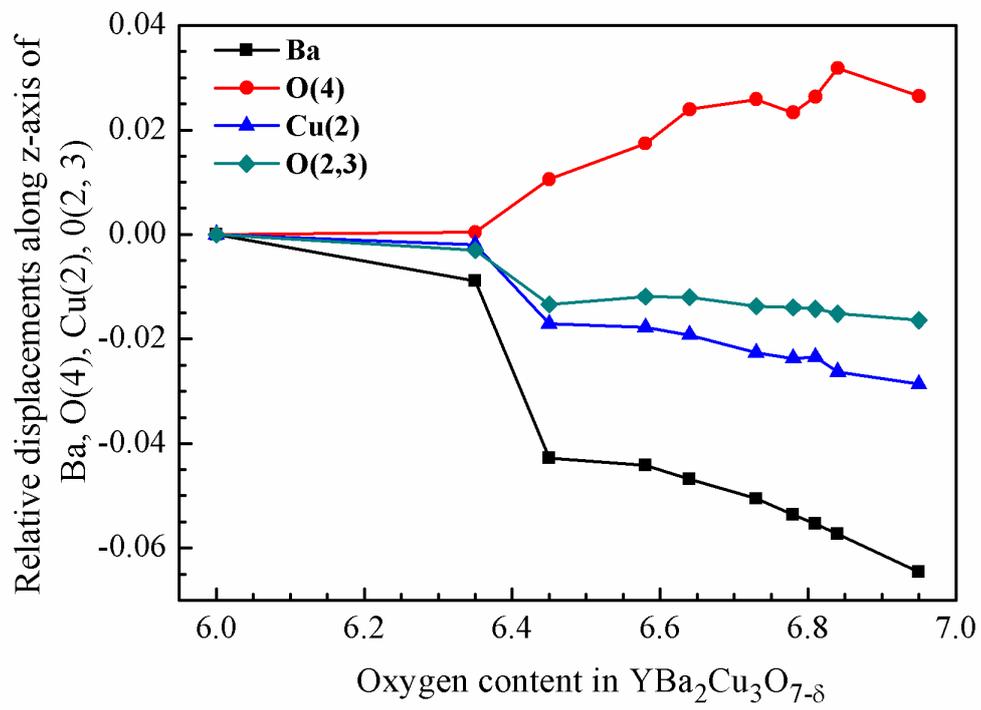

Figure. 4

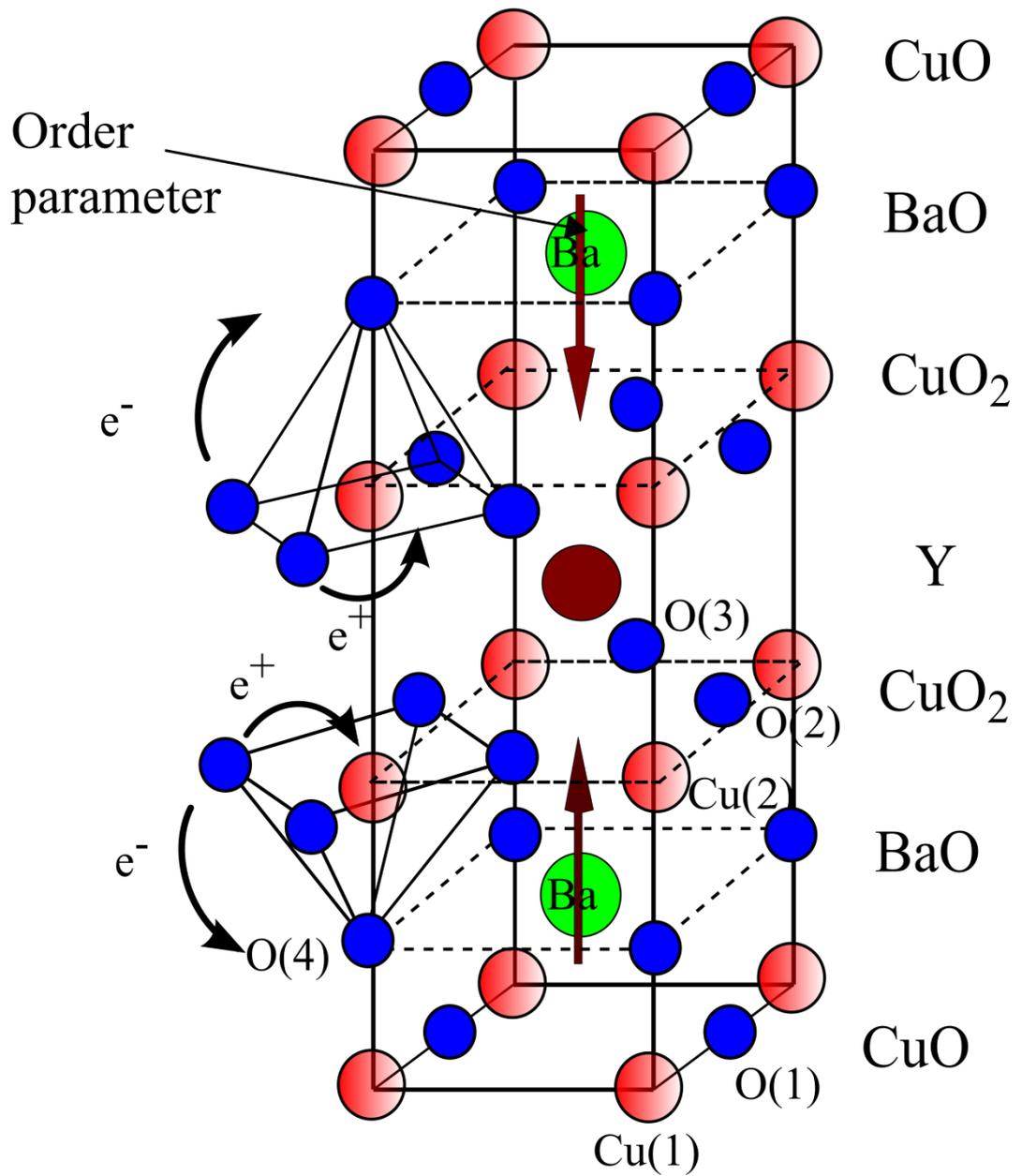

**Figure. 5**

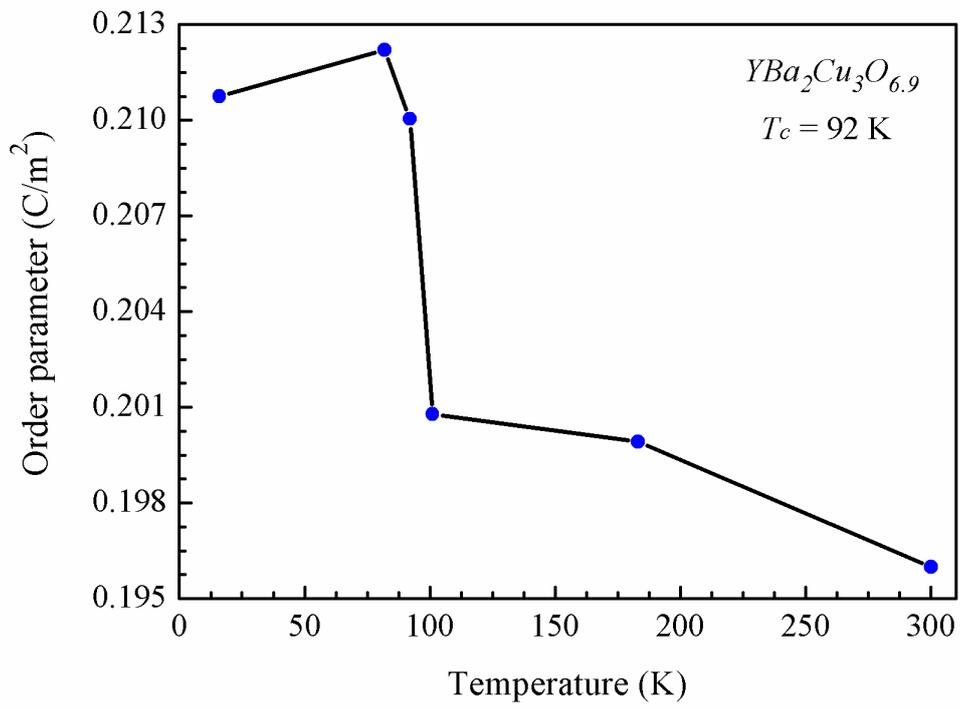

**Figure.6**

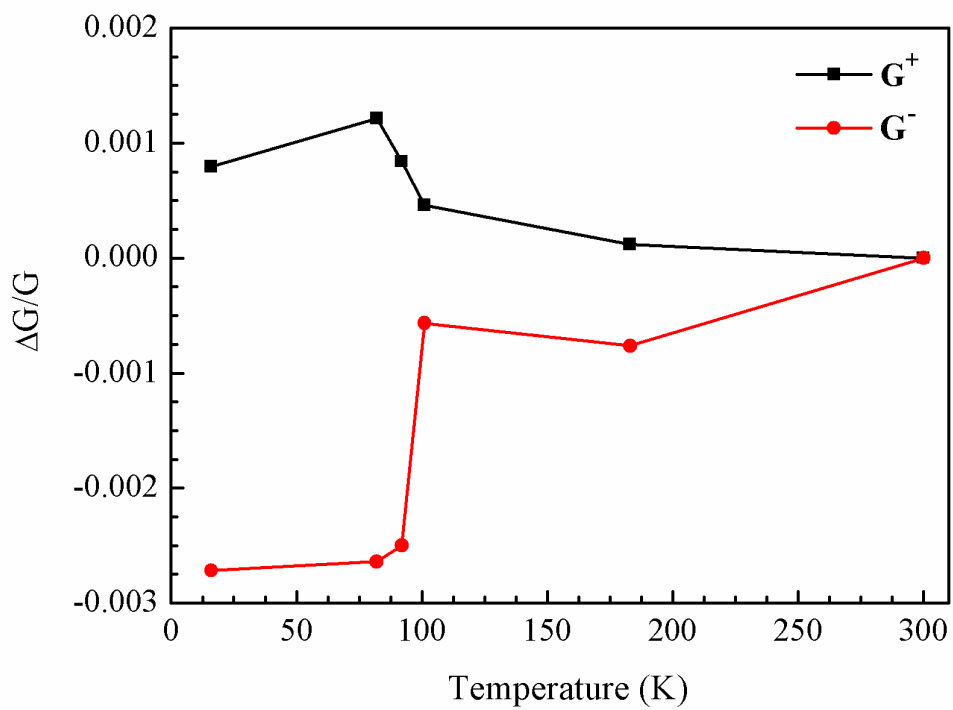

**Figure. 7**

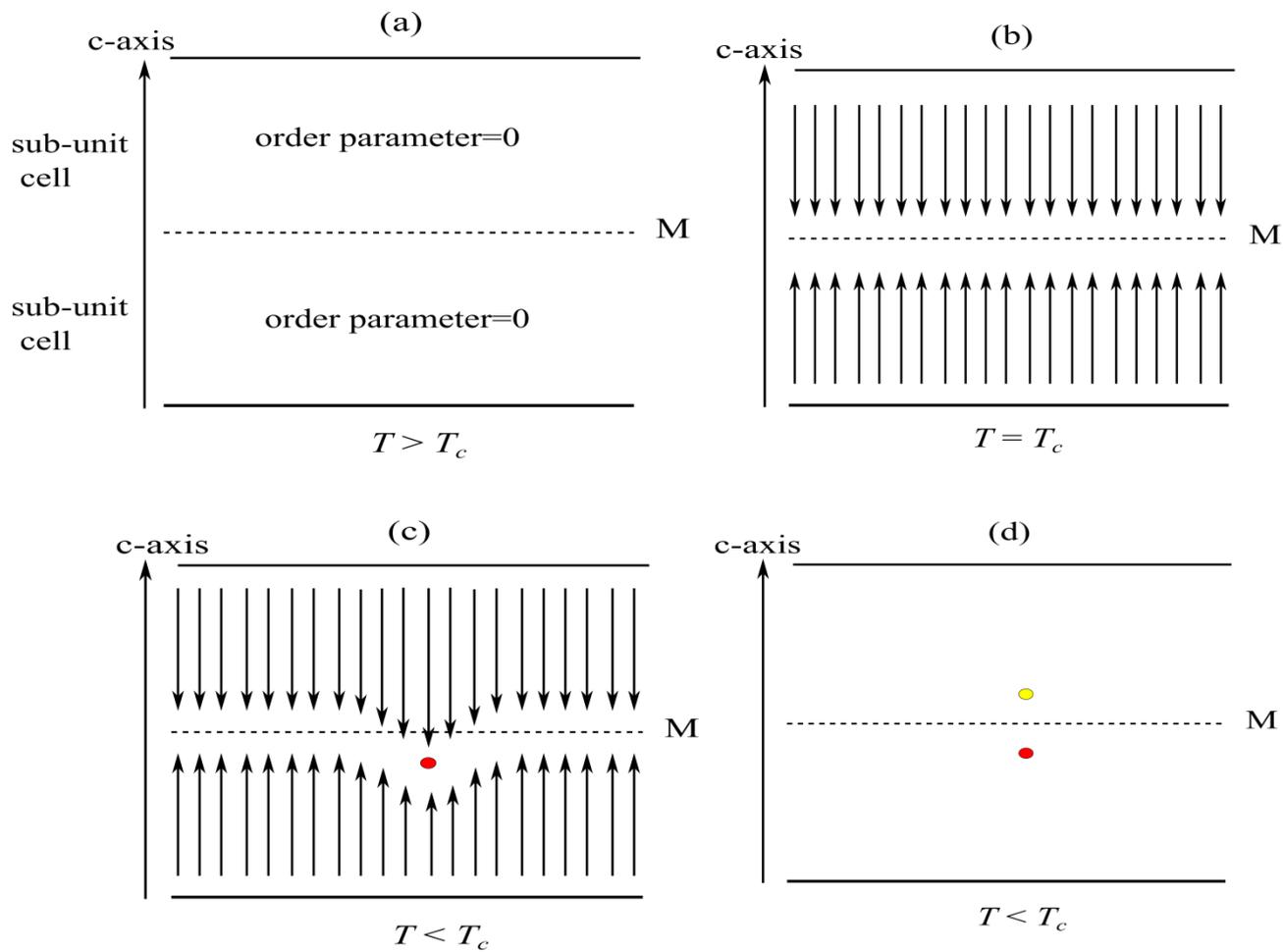

**Figure. 8**